\documentclass[referee]{raa}  

\usepackage{times}             
\usepackage{natbib}
\usepackage{amssymb,amsmath}
\usepackage[section]{placeins}
\bibpunct{(}{)}{;}{a}{}{,}
\usepackage{float}
\usepackage{graphicx}
\usepackage{hyperref}
\hypersetup{pdftitle = The title of my PDF, pdfauthor = My name, pdfsubject= The subject, pdfkeywords = keyword1 keyword2 keyword3} 
\hypersetup{colorlinks = true, linkcolor = green, anchorcolor = red, citecolor = blue, filecolor = red,  urlcolor = red}
\begin{document}

   \title{{ Timing analysis of the black hole candidate EXO~1846--031 with \emph{Insight}-HXMT monitoring}
}

   \volnopage{ {\bf 20XX} Vol.\ {\bf X} No. {\bf XX}, 000--000}      
   \setcounter{page}{1}          

   \author{
He--Xin Liu\inst{1,2*}, Yue Huang\inst{1,2*}, Guang--Cheng Xiao\inst{1,2}, Qing--Cui Bu\inst{1,3}, Jin--Lu Qu\inst{1,2}, Shu Zhang\inst{1},Shuang--Nan Zhang\inst{1,2}, Shu--Mei Jia\inst{1,2}, Fang-Jun Lu\inst{1}, Xiang Ma\inst{1}
% ...
}
   \institute{ Key Laboratory of Particle Astrophysics, Institute of High Energy Physics, Chinese Academy of Sciences, Beijing 100049, China; {\it liuhexin@ihep.ac.cn}
        \and
              University of Chinese Academy of Sciences, Chinese Academy of Sciences, Beijing 100049, China
	\and
Institut f\"ur Astronomie und Astrophysik, Kepler Center for Astro and Particle Physics, Eberhard Karls Universit\"at, 72076 T\"ubingen, Germany
\and 
 Department of Astronomy, Beijing Normal University, Beijing 100088, China
\and
 Department of Astronomy, Tsinghua University, Beijing 100084, China
\and
Department of Physics, Tsinghua University, Beijing 100084, China
\and
Department of Engineering Physics, Tsinghua University, Beijing 100084, China
\and 
Key Laboratory of Space Astronomy and Technology, National Astronomical Observatories, Chinese Academy of Sciences, Beijing 100012, China
\and
College of physics Sciences \& Technology, Hebei University, Baoding 071002, Hebei Province, China
\and 
Computing Division, Institute of High Energy Physics, Chinese Academy of Sciences, Beijing 100049, China
\and 
School of Physics and Optoelectronics, Xiangtan University, Xiangtan 411105, China 
\vs \no
   {\small Received 20XX Month Day; accepted 20XX Month Day}
}

\abstract{ We present the observational results from a detailed timing analysis of the black hole candidate EXO~1846-031 during its outburst in 2019 with the observations of \emph{Insight}-HXMT, \emph{NICER} and \emph{MAXI}. This outburst can be classified roughly into four different states. Type-C quasi-periodic oscillations (QPOs) observed by \emph{NICER} (about 0.1--6 Hz) and \emph{Insight}-HXMT (about 0.7--8 Hz) are also reported in this work. Meanwhile, we study various physical quantities related to QPO frequency. The QPO rms--frequency relationship in the energy band 1--10 keV indicates that there is a turning pointing in frequency around 2 Hz, which is similar to that of GRS~1915+105. A possible hypothesis for the relationship above may be related to the inclination of the source, which may require a high inclination to explain it. The relationships between QPO frequency and QPO rms, hardness, total fractional rms and count rate have also been found in other transient sources, which can indicate that the origin of type--C QPOs is non-thermal.
}

   \authorrunning{H. X. Liu, Y. Huang \& J. L. Qu  }            
   \titlerunning{Timing analysis of EXO~1846--031 }  

   \maketitle

%
%
\section{Introduction}           
\label{Introduction}
Black hole low mass X--ray binaries (BH--LMXBs), mostly  known as the transient systems (Black hole Transients, BHTs),  spend most of their lives in a quiescent state but outburst suddenly for maybe weeks or months. During the quiescent state of the BHTs, the luminosity is usually lower than $10^{33}$ erg/s, however, when it goes into an outburst, the source can reach a peak luminosity very fast, after which the luminosity will gradually decrease to the level of the quiescent state. During the outburst, BHTs usually exhibit four spectra states \citep{2005Ap&SS.300..107H}, namely the low-hard state (LHS), the hard-intermediate state (HIMS), the soft-intermediate state (SIMS), and the high-soft state (HSS). The X--ray observations indicate that the BHTs are in a hard state at the beginning of the outburst, and then will evolve into a soft state. Between the hard and soft states, two intermediate states (HIMS and SIMS) are defined based on their spectra and timing properties \citep{2005Ap&SS.300..107H}. The LHS mainly occurs in the early and later stages of the outburst, during which the energy spectra are dominated by a hard component, with a power law photon index varying between 1.5 and 1.7. The variability amplitude in LHS can reach a level higher than 20\%. In the HIMS, the variability gets weaker and the fractional rms usually fluctuates between 5--20\%. The SIMS exists for only a small fraction of time during the whole outburst, in which the spectra become softer, and the X-ray variation is weak (rms $\la 10\%$). When the source evolves into the HSS, the rms will drop to less than 3\%, and the energy spectra is dominated by the disc components. 

Low-frequency Quasi-periodic Oscillations (LFQPOs), which are defined as narrow peaks in the power spectra, have been observed in most BHTs (see \citealt{2006csxs.book...39V} for details). Three types of LFQPOs in BHTs with Fourier frequencies ranging from a few mHz to tens of Hz are classified based on their timing features (fractional rms, quality factor $Q =  \nu_0/(2\Delta)$, where $\nu_0$ is the center frequency, and $\Delta$ is the half width at half maximum) in the power density spectra (PDS), noise component, and phase lag properties \citep{1999ApJ...514..939W,2002ApJ...564..962R,2005ApJ...629..403C,2011MNRAS.418.2292M}. Type--A QPOs ususlly appear in the soft state of an outburst when the light curve has rms $\la 3\%$. The PDS of type--A QPOs are characterized by low-amplitude, red noise and the absence of harmonics. Type--B QPOs appear in the soft-intermediate state when the rms of light curve is 3\% -- 5\% \citep{2015ApJ...808..144K}, whose PDS are characterized by weak harmonic and power law noise. Generally seen in LHS and HIMS, type-C QPOs are characterized as strong narrow peak and associated with strong `flat-top' noise, harmonics and possible sub harmonic in the PDS. During the hard states (LHS and HIMS), the centroid frequencies of type--C QPOs regularly increase with source luminosity. 

The origin of the type--C QPOs has been studied extensively. Theoretical models that explain the origin of type--C QPO can be divided into two categories: the instability of accretion flows \citep{1999ApJ...525L.129T,2004ApJ...612..988T,2012A&A...545A..40V} and the geometric effects of general relativity \citep{1998ApJ...492L..59S,1999NuPhS..69..135S,2006ApJ...642..420S,2009MNRAS.397L.101I}. Considering the instability, physical models such as transition layer \citep{2004AIPC..714..383T}, corona oscillation \citep{2010MNRAS.404..738C} and the Accretion-Ejection Instability (AEI) mechanism \citep{2003astro.ph.10026V} have been proposed in recent years. The geometric effect involved is mainly a relativistic precession model, which was first proposed by \citet{1998ApJ...492L..59S}, then modified and improved by \citet{2006ApJ...642..420S}. This model can explain the main observational characteristics of LFQPOs, for example, the frequency evolution of LFQPOs, energy spectra and its phase-resolved spectra. In recent years, more and more observations by X--ray telescopes support that the type--C QPOs may have a geometric origin, with Lense-Thirring precession of the entire inner flow \citep{2009MNRAS.397L.101I,2015MNRAS.448.1298P,2017MNRAS.464.2643V}. 

In 1985, the European X-ray Observatory Satellite (\emph{EXOSAT}) discovered the black hole X--ray binary candidate EXO~1846-031. By analyzing data from the first outburst in 1985, \citet{1993A&A...279..179P} proposed that the center of the source is most likely a black hole. The source erupted again and was detected by the Monitor of All-sky X-ray Image (\emph{MAXI}) on July 23, 2019 and the 4--10 keV flux reached about 100 mCrab on July 28 \citep{2019ATel12968....1N}. \emph{Insight}--HXMT began its observation from August 2, 2019, and the preliminary energy spectra analysis shows that the X-ray spectral properties are typical for outbursts in BH--LMXBs \citep{2019ATel13036....1Y}. The Neutron Star Interior Composition Explorer (\emph{NICER}) observed EXO~1846-031 between 2019 July 31 and 2019 October 25 \citep{2019ATel12976....1B}.

In this work, we use the observations by \emph{Insight}-HXMT, \emph{NICER} and \emph{MAXI} to study the temporal features of this source. Section \ref{sec:2} presents the introduction of observations and data reduction. In section~\ref{sec:3}, the timing results are described in detail. We make a summary and some discussions of our results in section \ref{sec:4}.


\section{Observations and Data analysis}
\label{sec:2}

\subsection{Observations}
\label{sec:2.1} 
The first Chinese X-ray astronomical satellite, the Hard X-ray Modulation Telescope (known as \emph{Insight}-HXMT) with three main instruments, was launched on June 15, 2017. More details about \emph{Insight}--HXMT can be found in \citet{2020SCPMA..63x9502Z}. The three instruments are the High Energy X-ray Telescope (HE), the Medium Energy X-ray Telescope (ME), and the Low Energy X-ray Telescope (LE). HE \citep{2020SCPMA..63x9503L}, with a total detection area of about 5000 cm$^{2}$, is comprised of 18 cylindrical NaI/CsI phoswich detectors in the 20--250 keV energy band. ME \citep{2020SCPMA..63x9504C}, which is made up of three boxes, contains 1728 Si-PIN detector pixels, with a total detection area of about 952 cm$^{2}$ in the 5--30 keV energy band. Using Swept Charge Device (SCD), LE \citep{2020SCPMA..63x9505C} is sensitive in 1--15 keV and its detection area is 384 cm$^{2}$. They all have collimators with large and small fields of view (FOVs) in their detectors. In this work, we only use the small FOVs of HE, ME and LE. Detailed discussions of its calibrations and background are given in \citet{2020arXiv200306998L} and \citet{2020arXiv200401432L} (HE), \citet{2020arXiv200306260G} (ME), \citet{2020arXiv200401432L} (LE), respectively. \emph{Insight}-HXMT began high--cadence pointing observations of EXO~1846-031 (Target of Opportunity) on August 2, 2019, and ended on October 25, 2019. 

Moreover, data from \emph{NICER} and \emph{MAXI} also provides meaningful results. \emph{NICER}'s X--ray Timing Instrument (XTI) is exceptionally sensitive to the energy ranging from 0.2 keV to 12 keV \citep{2014SPIE.9144E..20A,2012SPIE.8453E..18P}. As for the \emph{MAXI}, one of its main purposes is to monitor the known X--ray sources for their intensity fluctuations over long periods of time by scanning the entire sky in both soft and hard X--rays. The probe energy segment of the Gas Slit Camera (\emph{MAXI}/GSC) is 2--30 keV \citep{2014SPIE.9144E..1OM}. This full outburst is observed by \emph{MAXI}.


\subsection{Data reduction}
\label{sec:2.2} 
Using the HXMT Data Analysis software (HXMTDAS) v2.02\footnote[1]{http://www.hxmt.org/index.php/usersp/dataan/fxwd}, we process the data acquired by \emph{Insight}--HXMT for 2019 outburst of EXO~1846--031 in five steps. Firstly, we calculate Pulse Invariant (PI) from the raw values of Pulse Height Amplitude (PHA) of each event, accounting for temporal changes in gain and energy offset. Secondly, Good Time Intervals (GTIs) are calculated for each instrument with particular criteria. The third step is reconstructing the split events for LE; as for ME, calculating the grade and dead time. Then, selecting the events with GTIs is the fourth step. Finally, from the screened events, respectively for the three instruments, scientific products can be extracted, including the light curves, energy spectra and backgrounds files. 

In this paper, we filter the data with the following criteria \citep{2018ApJ...866..122H}: (1) pointing offset angle $ < 0.04^\circ$; (2) elevation angle $ > 6^\circ$ ; (3) value of the geomagnetic cutoff rigidity $> 8$. However, the detectors of the LE are sensitive to bright earth, so we set the bright earth elevation angle (DAY\_ELV) $ > 30^\circ$, while for the ME and HE DAY\_ELV $> 0^\circ$. 

Using the steps above, the light curves with $2^{-5}$ s time resolution and background light curves are generated from LE, ME and HE in energy bands: 1--10 keV, 10--20 keV and 25--150 keV respectively.

The light curves with  $2^{-5}$ s time resolution of \emph{NICER} are generated with cleaned events data by software XSELECT. The data of \emph{MAXI} are downloaded from the official website\footnote[2]{http://maxi.riken.jp/top/index.html}.

\subsection{Data Analysis and Methods}
\label{sec:2.3} 
The hardness-intensity diagram (HID) and hardness-rms diagram (HRD) are conventional methods to analyzing the spectral properties of X-ray binary transients. In many cases, the shape of the HID from BHTs tracks a like `q' pattern, if the source performs a complete outburst. We use timing properties and HID of this source to distinguish spectral states. The hardness ratio is defined as the ratio of count rate in 4--10 keV and 2--4 keV for LE \& \emph{MAXI}/GSC in this work.

In order to analyze the timing properties, we use the POWSPEC to create power density spectrum (PDS) from each light curve by rms  normalization \citep{1991ApJ...383..784M} and split the extracted light curves into 64 s segments including 2048 bins.
We fit the  the PDS using XSPEC V12.10 for different components. For the possible QPO components, we fit the poisson-extracted PDS using a model consisting of several Lorentzians accounting for the possible broad band noise (BBN), the possible QPO fundamental and (sub) harmonics.
Since we use the rms normalization in PDS, the fractional rms of QPO is given by \citep{2015ApJ...799....2B}:
\begin{equation}
   rms_{\rm QPO}=\sqrt{R}\times \frac{(S+B)}{S}.	\label{eq:quadratic}
\end{equation}
Here \emph{R} is the normalization of the Lorentzian component of the QPO, \emph{S} is the source count rate, while \emph{B} is the background count rate.

\section{Results}
\label{sec:3}
\subsection{Hardness and Timing Evolution}
\label{sec:3.1} 
\begin{figure}
	\centering
	\includegraphics[width=14cm]{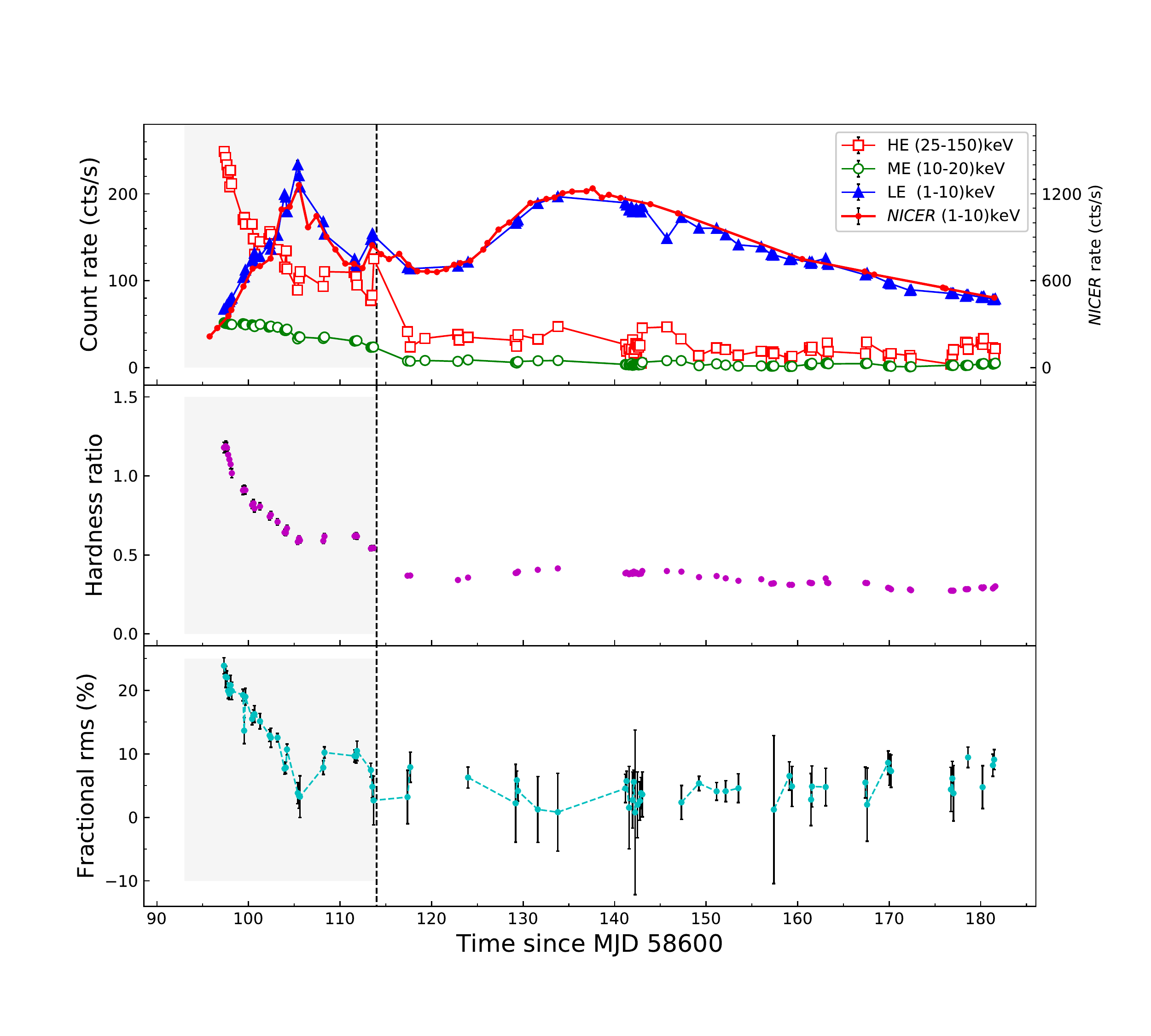}
    \caption{Light curves, hardness ratio and total fraction rms evolution for EXO~1846--031 during its 2019 outburst observed by \emph{Insight}--HXMT and \emph{NICER}. The top panel presents the light curves by three energy bands: 1--10 keV (LE, blue triangles; \emph{NICER}, red points),10--20 keV (ME, green circles), and 25--150 keV (HE, red squares). The middle panel shows the hardness between 4--10 keV (LE) and 2--4 keV (LE) count rate. We show the total fraction rms of the LE (1--10 keV) light curve in the bottom panel, where the frequency ranges from $2^{-5}$ Hz to 32 Hz. Type--C QPOs are detected from MJD 58697 to MJD 58715, of which observations are shown in gray areas. Each point corresponds to one observation of \emph{Insight}--HXMT or \emph{NICER}. }
    \label{fig:LC}
\end{figure}

Based on the \emph{Insight}--HXMT data, we show light curves, hardness, and total fractional rms of EXO~1846--031 during this outburst in Figure \ref{fig:LC}. Throughout the outburst, the light curves of LE and \emph{NICER} show similar behaviors, which can be seen in the top panel of Figure \ref{fig:LC}. The count rate in low energy (LE 1--10 keV) increases rapidly during the rise phase until it reached its peak at 233.39 cts/s on MJD 58705. As for the high energy (HE 25--150 keV) and medium energy (ME 10--20 keV), the count rate decreases monotonously from 248.85 cts/s, 51.70 cts/s to a constant level, respectively. In the middle panel of Figure \ref{fig:LC}, the hardness decreases sharply at first. After reaching a small peak, the hardness begins to fall again to a stable low level. Correspondingly, the count rate of LE (1--10 keV) also has a small fluctuation. As can be seen from the fractional rms evolution diagram (the bottom panel of Figure \ref{fig:LC}), most type-C QPOs (section \ref{sec:3.2}) appear in fractional rms $ > 5\%$, which is consistent with typical black hole transient sources, for example, H1743--322 \citep{2013MNRAS.431.2285Z}. 

\begin{figure}
    \centering
    \includegraphics[width=12cm]{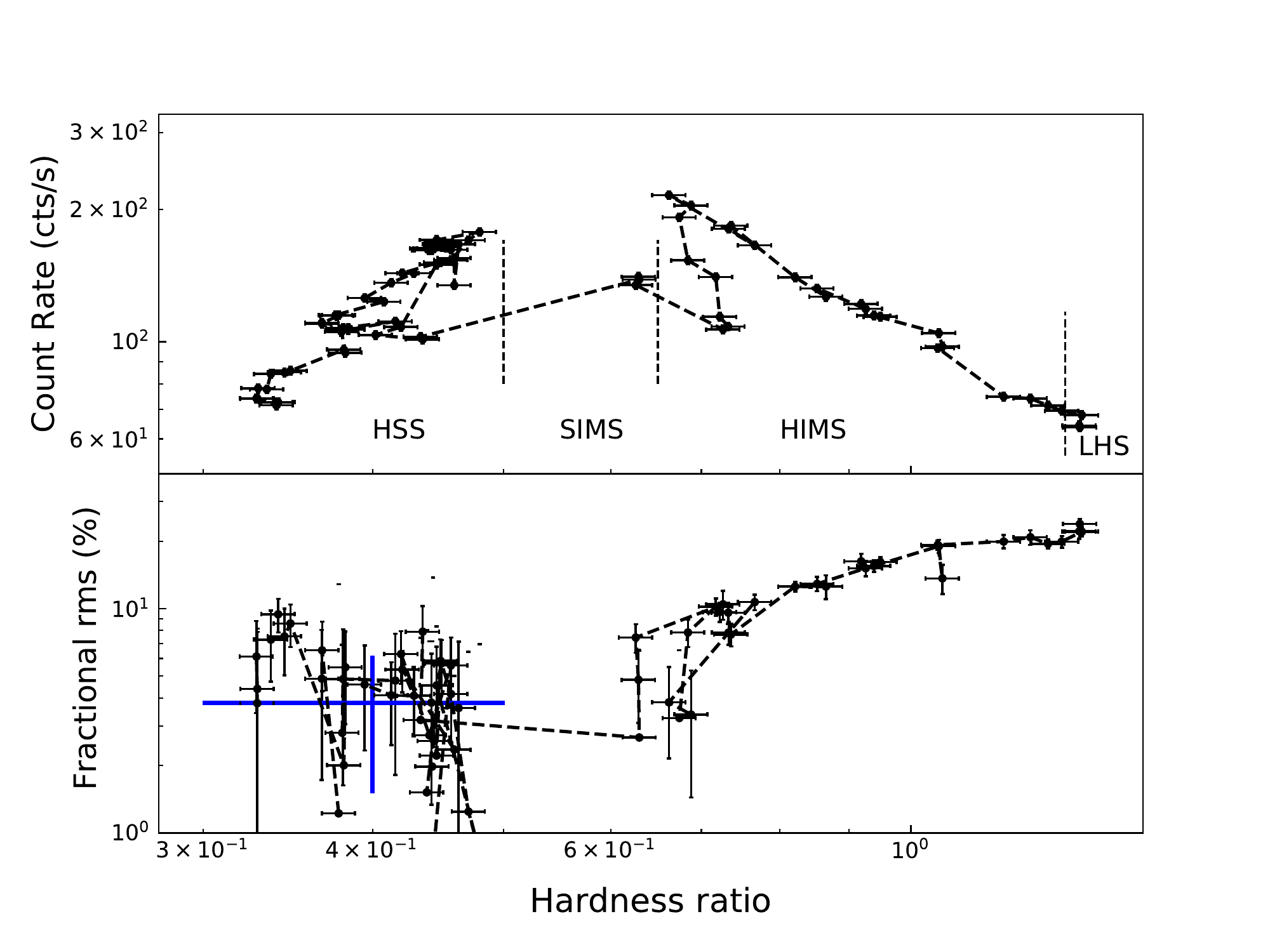}
    \caption{The HRD and HID of EXO~1846--031 in this outburst are presented in the top and bottom panels respectively. The fractional rms for 1--10 keV (\emph{Insight}--HXMT/LE) are integrated within $2^{-5}$--32 Hz. The hardness is  defined as the ratio between LE's 4--10 keV and 2--4 keV count rate. The blue line represents the average of the fractional rms in HSS.}
    \label{fig:HRD and HID}
\end{figure}

\begin{figure}
    \centering
    \includegraphics[width=12cm]{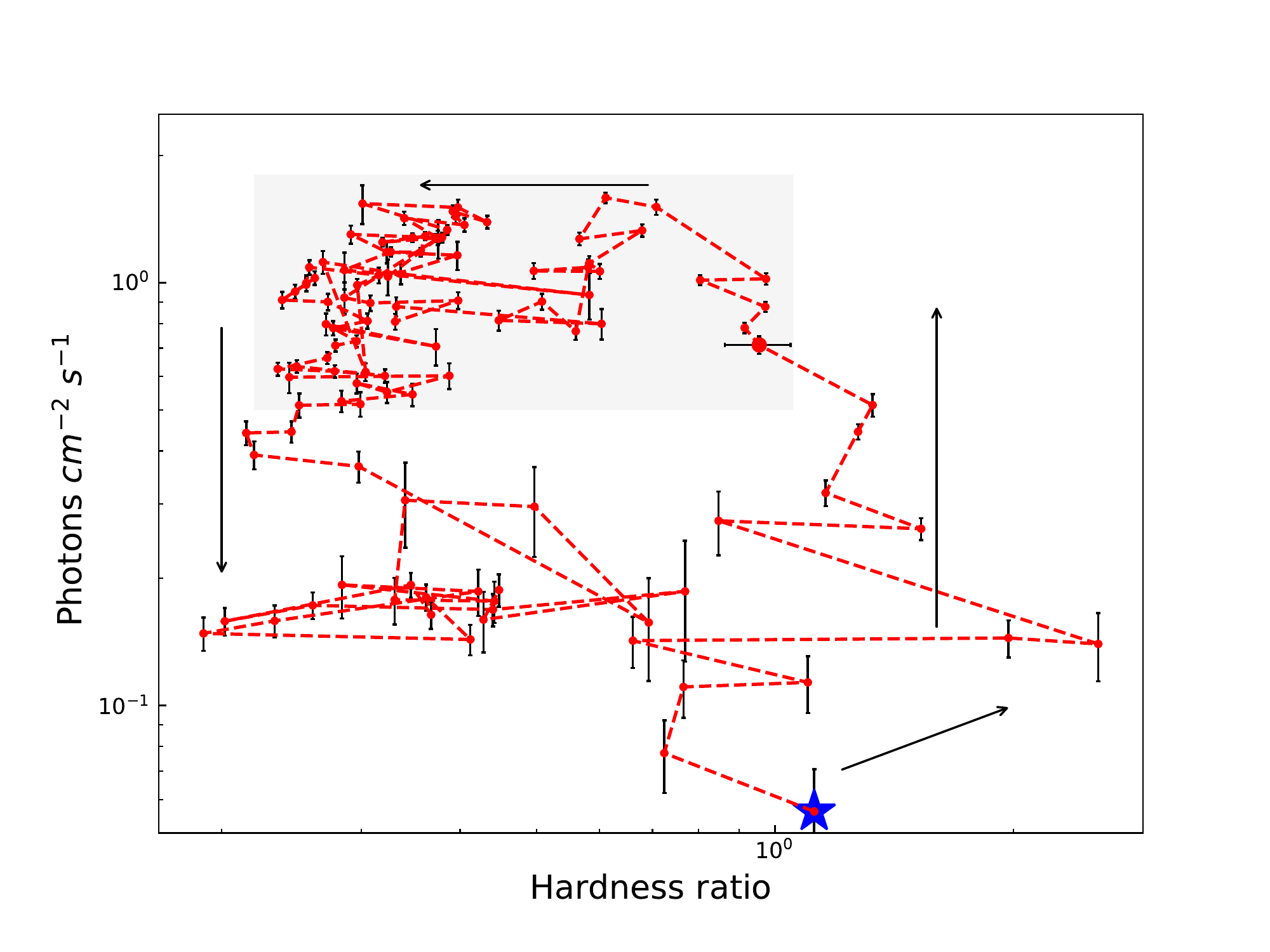}
    \caption{The evolution of its count rate with hardness during this complete outburst of EXO~1846--031 using \emph{MAXI} GSC data. The hardness is the ratio of 4--10 keV to 2--4 keV energy band count rate, and the vertical axis represents the count rate of 2--10 kev energy segment. The grey area corresponds to the period when the \emph{Insight}--HXMT observations were performed. The blue star represents the first observation.}
    \label{fig:MAXI HID}
\end{figure}

In Figure \ref{fig:HRD and HID}, we present the HID and HRD of EXO~1846-031 using \emph{Insight}--HXMT/LE count rate. The results show that no complete HID shape is revealed, because the decrease phase of the outburst is not observed. Therefore, we use data from the longer observation period of \emph{MAXI} to draw more complete HID in Figure \ref{fig:MAXI HID}, where a complete `q' shape is shown. Comparing Figure \ref{fig:MAXI HID} with Figure \ref{fig:HRD and HID}, the observations by LE \& \emph{NICER} only cover a part of the complete `q' pattern due to low count rate. Throughout the outburst, the hardness decreases from $\sim$ 1.3 to $\sim$ 0.3 and the fractional rms of the continuum's PDS decreases from $\sim$ 25\% to $\sim$ 3\%, respectively, which indicates that the source undergoes state transitions. Moreover, using QPOs positions and spectral evolution in HID, we can identify the different spectral states of the outburst, i.e., LHS, HIMS, SIMS, and HSS. 

For the first three observations of \emph{Insight}--HXMT, the X-ray flux increases but hardness ratio barely changes. At the same time, we compare the position of these three observations in HID of \emph{MAXI}, which is consistent with the phenomena that X-ray flux increases rapidly while hardness ratio remains unchanged. Those characteristics show that the source may be in its LHS. During the LHS, the frequency of QPOs increases from 0.71 Hz to 0.96 Hz, and the total fractional rms decreases from $\sim$ 23.87\% to $\sim$ 22.03\%. Following the LHS, the count rate in low energy enhances from 67.31 cts/s and gradually reaches its peak at 233.39 cts/s with the hardness ratio decreasing from 1.3 to 0.7 (see Figure \ref{fig:HRD and HID}). The frequency of QPOs increases from 0.96 Hz to 6.59 Hz (see Figure \ref{fig:QPO frequency and LW}), and the total fractional rms decreases from $\sim$ 22\% to $\sim$ 10\%. The evolution of the hardness ratio and the timing properties show that the source goes into its HIMS. After the flux peak, the X-ray flux of the source drops steeply to 117.47 cts/s with an constant hardness ratio. Those characteristics seem to be consistent with the LHS at rise phase of the outburst. However, during this phase, the near constant frequency of the QPOs and the stable total fractional rms indicate that the source is in the HIMS. On about MJD 58713, the outburst enters into the second hump, the peak value of the LE count rate reaches about 154.16 cts/s. During this hump, the total fractional rms drop to $\sim$ 3\% and the hardness is less than 0.6. Those phenomena are consistent with SIMS, although no type--B QPO is observed. When the source leaves the SIMS, the evolution of the light curve enters the last count hump. The results show that the hardness is less than 0.3, and the total fractional rms remains near a constant value. The outburst entered the HSS. The rough classification between the four states is marked by vertical dashed lines in Figure \ref{fig:HRD and HID}, while the detailed state analysis is required in combination with the energy spectrum (Ren et al., manuscript in preparation).


\begin{figure*}
    \centering
    \subcaptionbox{MJD 58697 (LE)}{
    \includegraphics[width=0.42\linewidth]{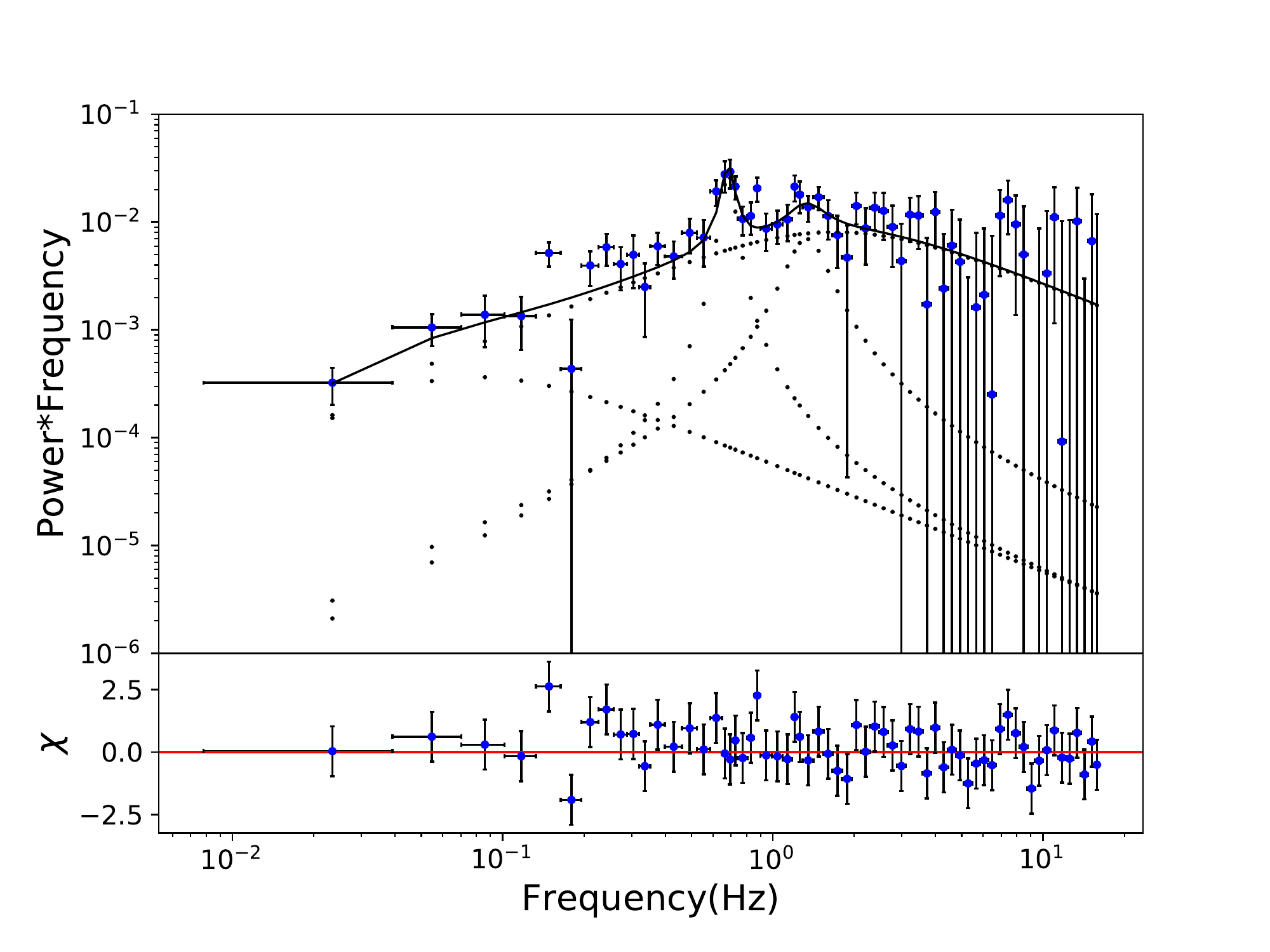}
    }
    \quad
    \subcaptionbox{MJD 58699 (LE)}{
    \includegraphics[width=0.42\linewidth]{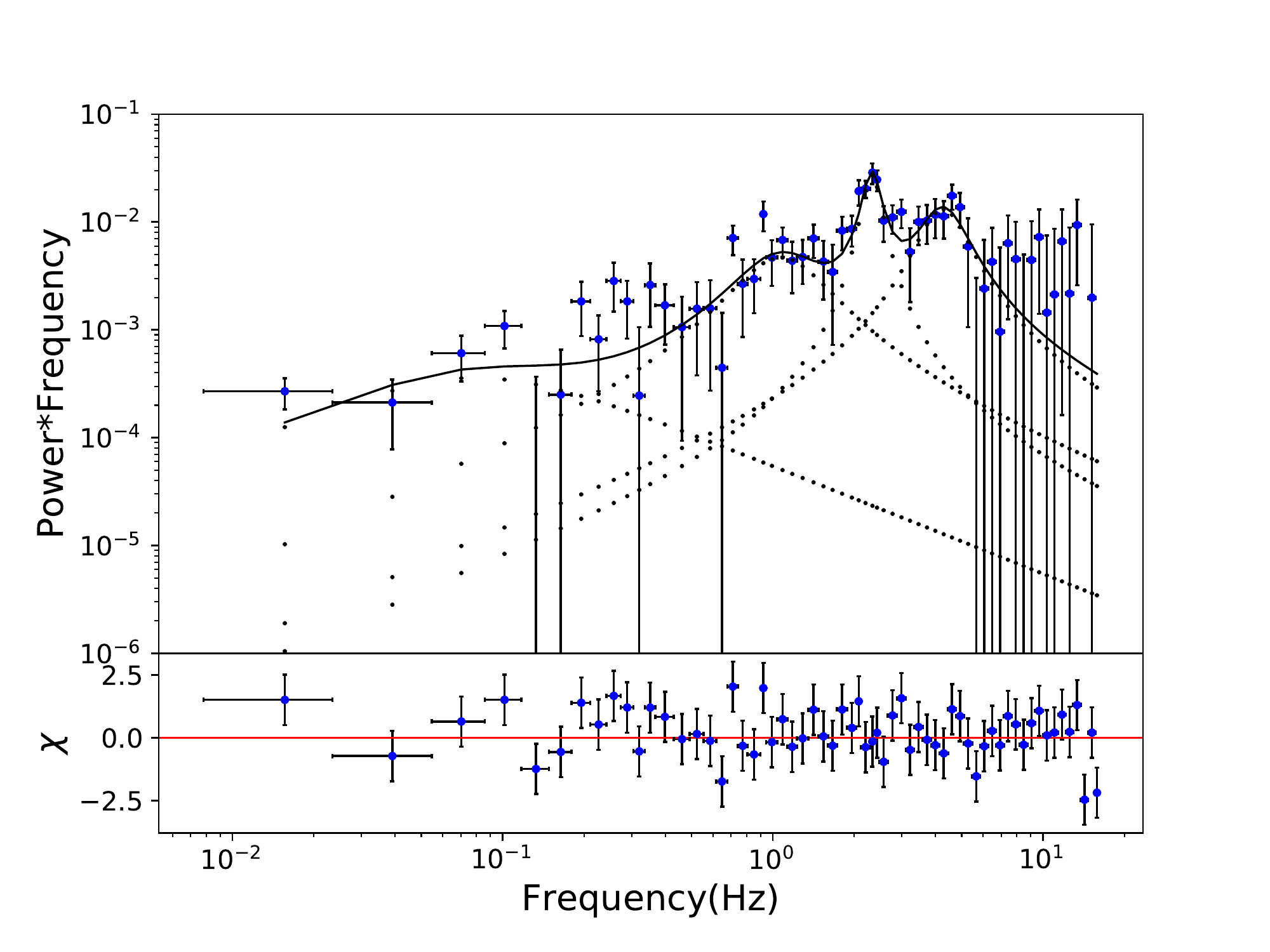}
    }
    \quad
    \subcaptionbox{MJD 58697 (ME)}{
    \includegraphics[width=0.42\linewidth]{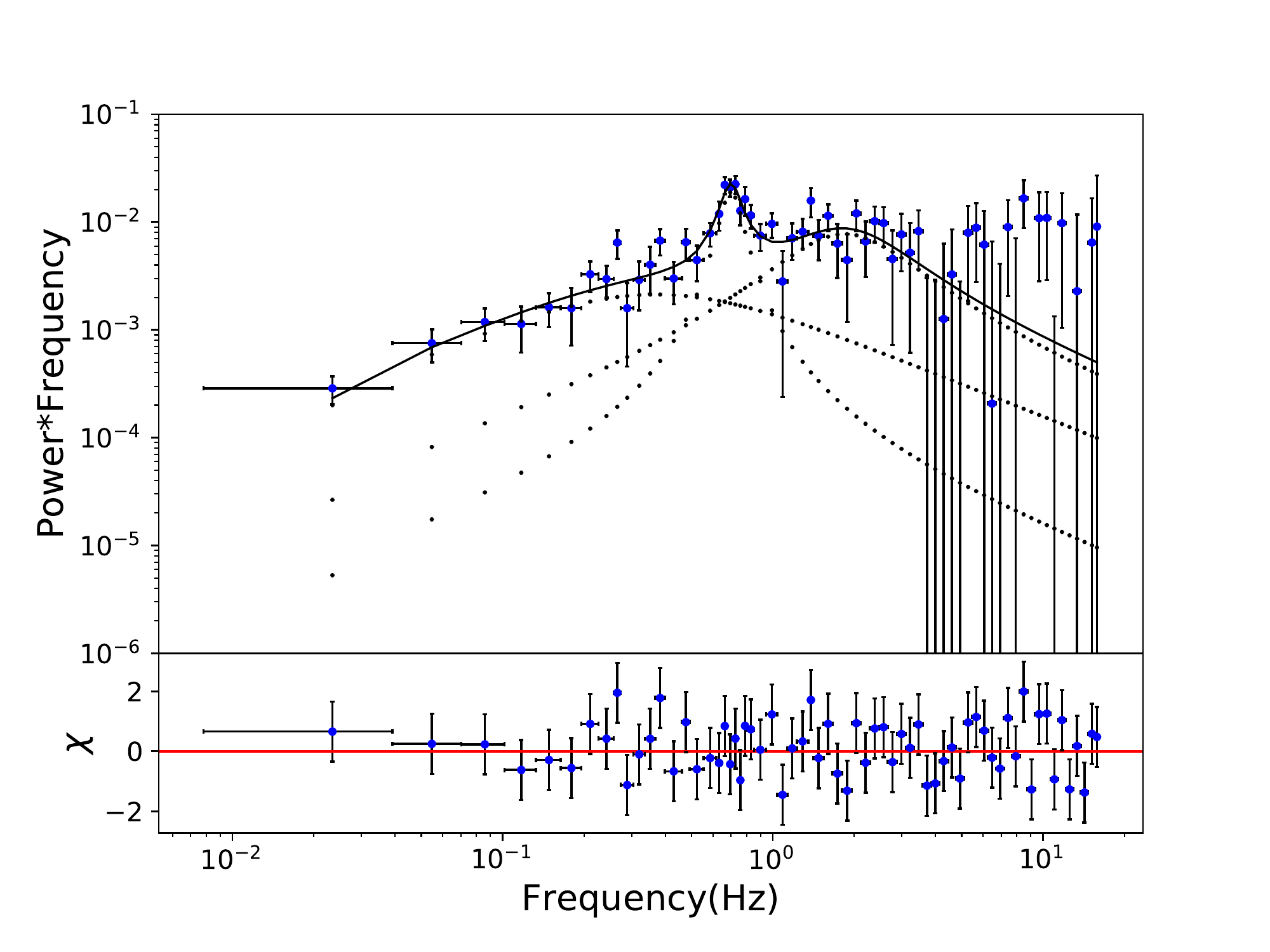}
    }
    \quad
    \subcaptionbox{MJD 58699 (ME)}{
    \includegraphics[width=0.42\linewidth]{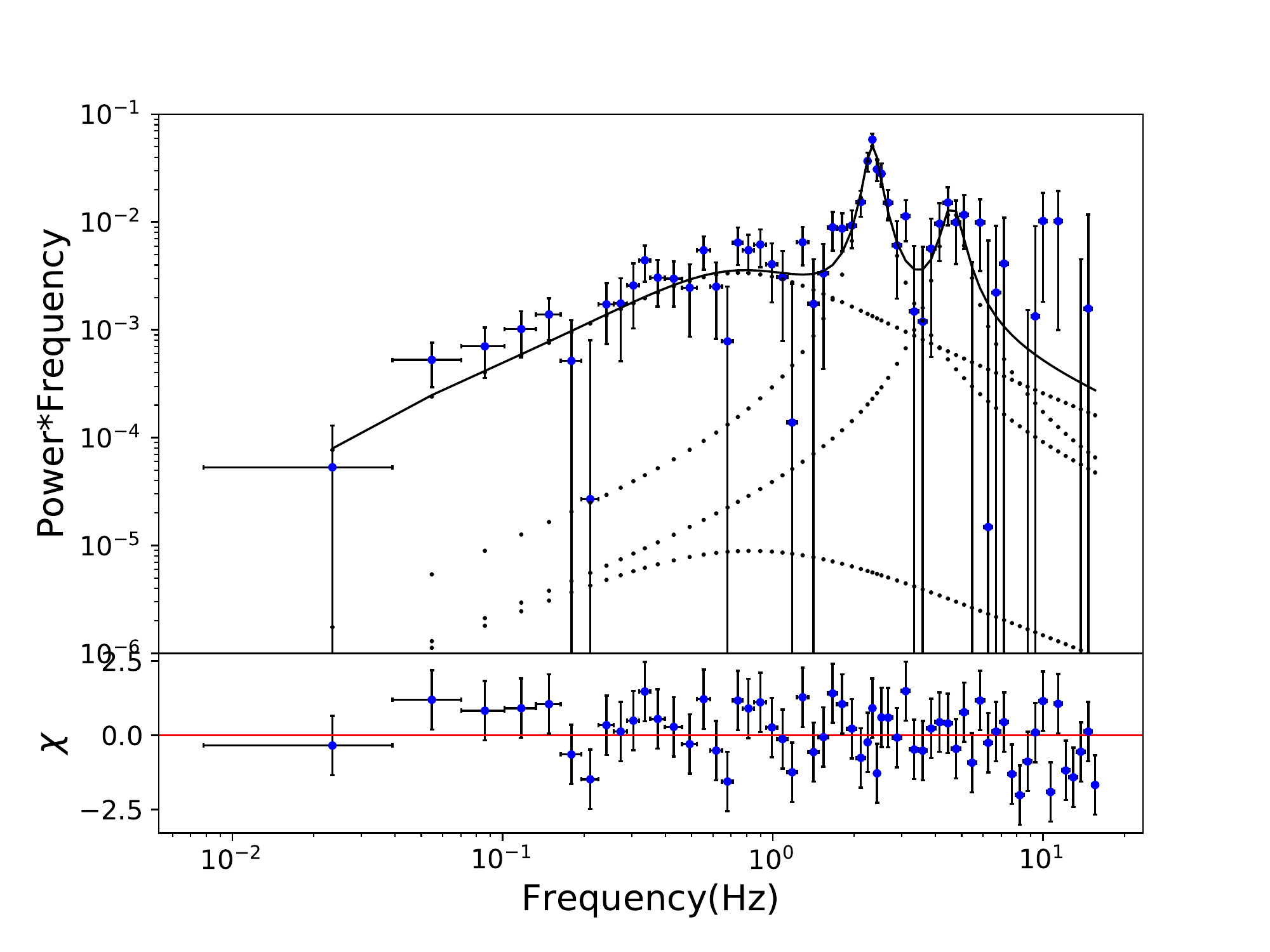}
    }
    \quad
    \subcaptionbox{MJD 58697 (HE)}{
    \includegraphics[width=0.42\linewidth]{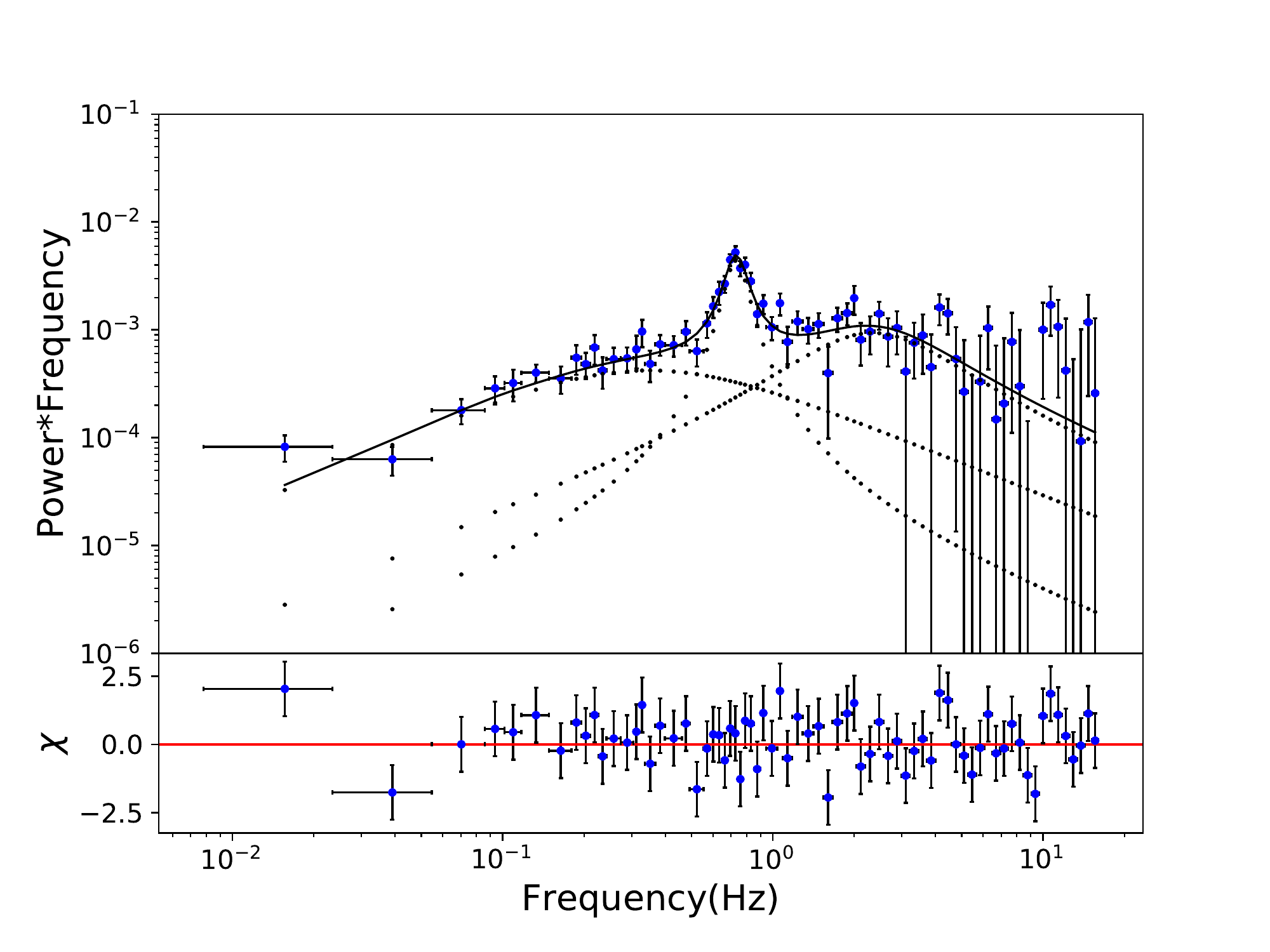}
    }
    \quad
    \subcaptionbox{MJD 58699 (HE)}{
    \includegraphics[width=0.42\linewidth]{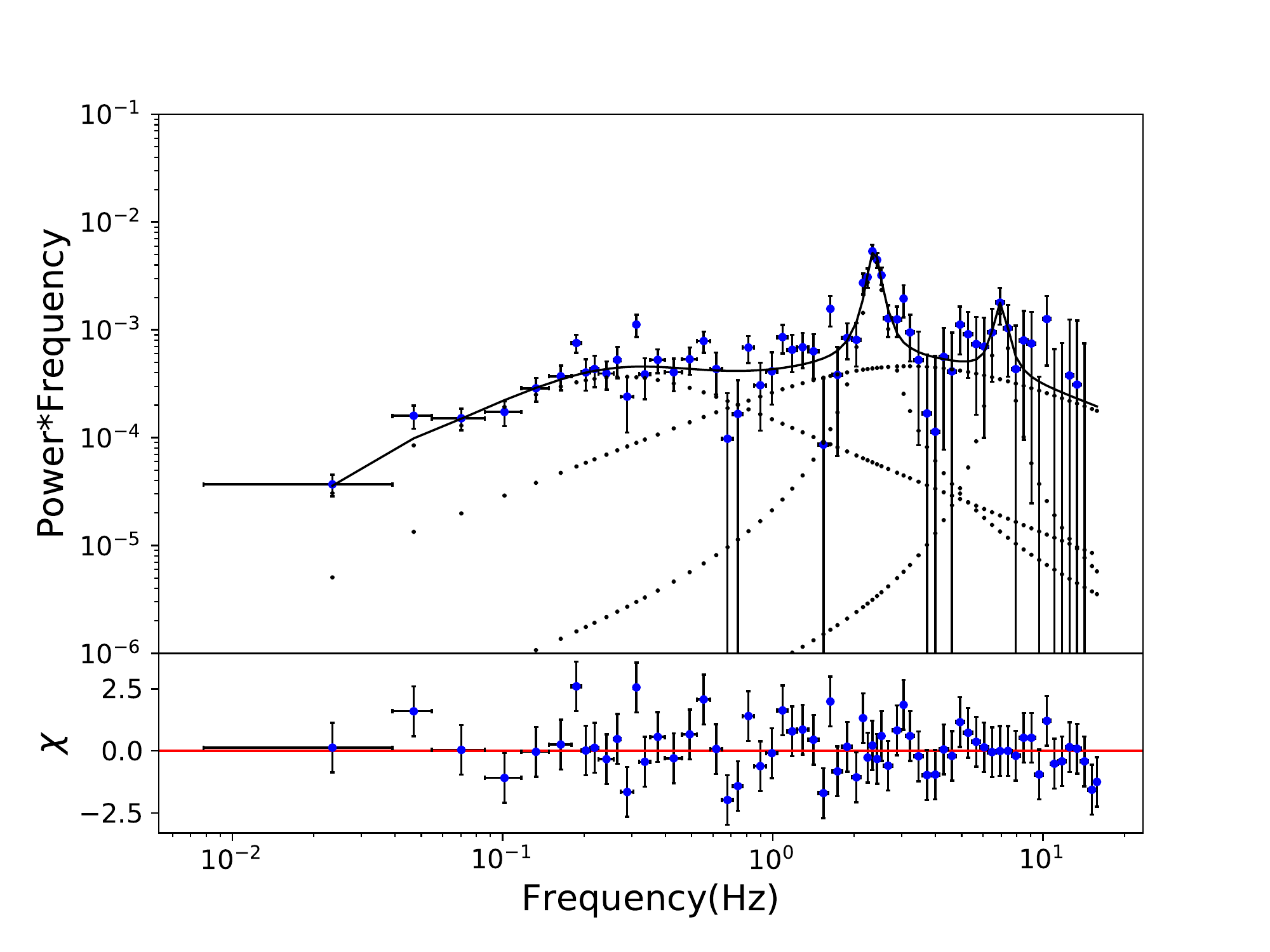}
    }
    \subcaptionbox{MJD 58695 (\emph{NICER})}{
    \includegraphics[width=0.42\linewidth]{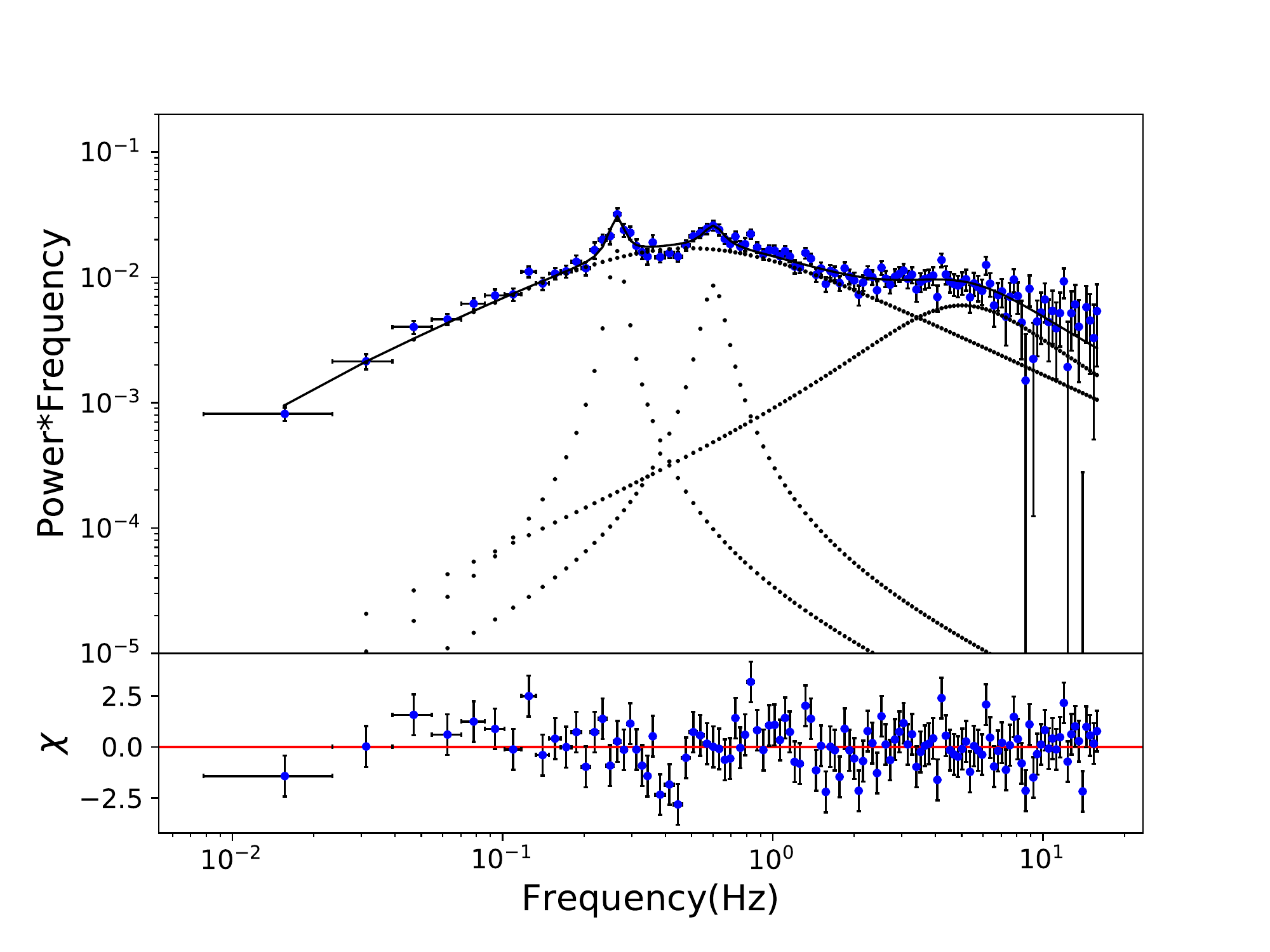}
    }
    \subcaptionbox{MJD 58699 (\emph{NICER})}{
    \includegraphics[width=0.42\linewidth]{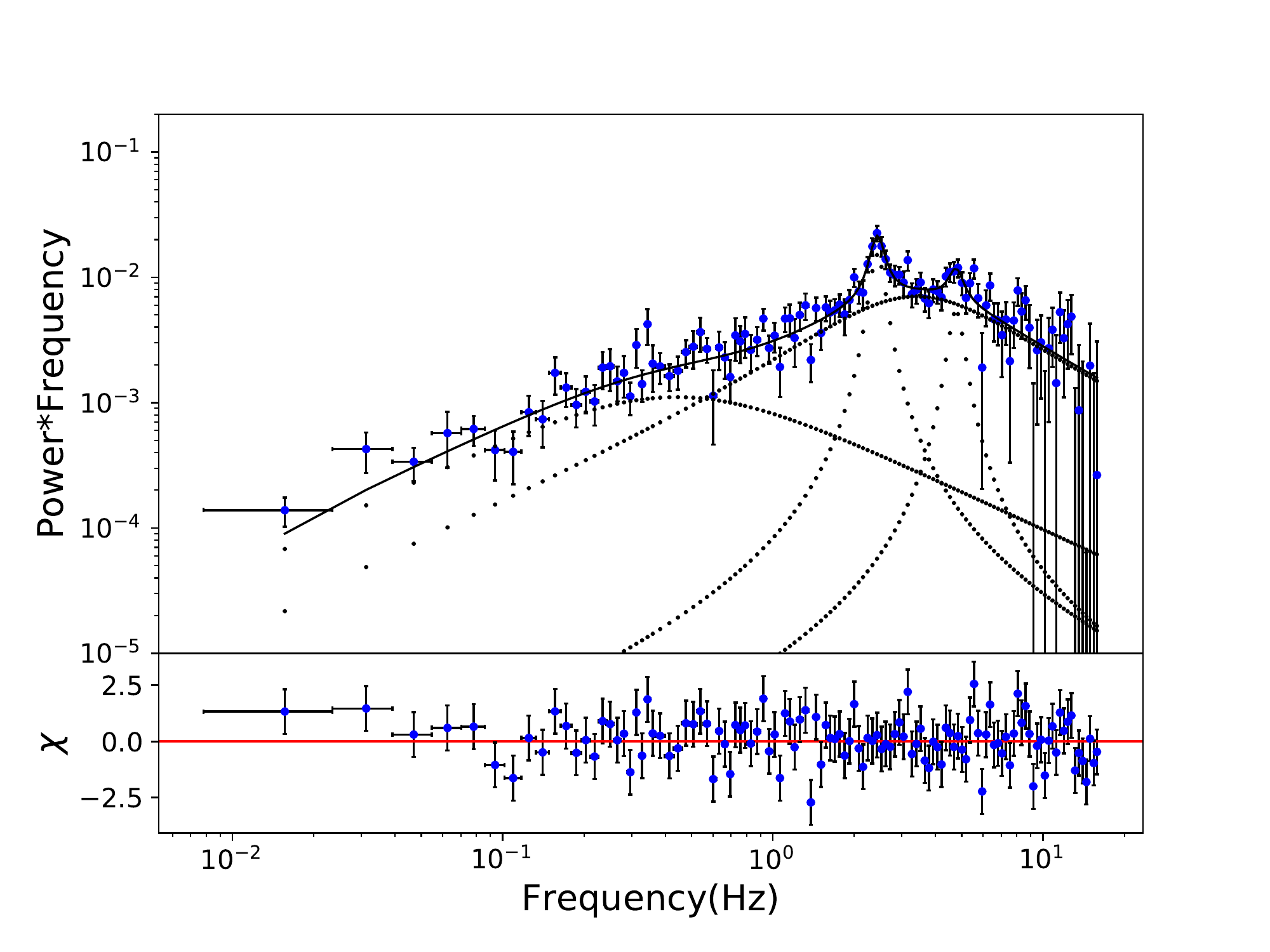}
    }

    \caption{Eight representative examples of QPO PDS. The top six panels are computed from \emph{Insight}-HXMT three instruments (LE:1--10 keV; ME:10--20 keV; HE:25--150 keV), corresponding to observations P021405000101 (left) and P021405000201 (right). PDS of two \emph{NICER} observations (2200760101, 2200760105) are shown on the bottom two panels.}
    \label{fig:PDS}
\end{figure*}

\subsection{Low-frequency QPOs}
\label{sec:3.2} 
To study the fast time-variability properties of EXO~1846--031, the PDS is calculated for each observation. The PDS shows a broad noise component and one or two narrow peaks at different frequency. It can be seen from the PDS diagram that the observed QPOs can be classified as type--C QPOs, and no type--B QPO is detected throughout the outburst. The QPOs are observed since the first day (MJD 58697) of the \emph{Insight}--HXMT observations and last for about 8 days. The \emph{NICER} observations started from MJD 58695 and the QPO observations last for 19 days. Figure \ref{fig:PDS} shows 8 representative examples (one representative observation at LHS and one at HIMS) and the fitting results of PDS for observations with different QPO frequencies. Using the fitting model mentioned in section \ref{sec:2.3}, we fit all the parameters of the QPOs as shown in Tables \ref{tab:LE QPOs}, \ref{tab:ME QPOs}, \ref{tab:HE QPOs} and \ref{tab:NICER QPOs}. The frequency range of the total QPOs observed is between 0.26 Hz and 8.42 Hz.

\begin{figure}
    \centering
    \includegraphics[width=12cm]{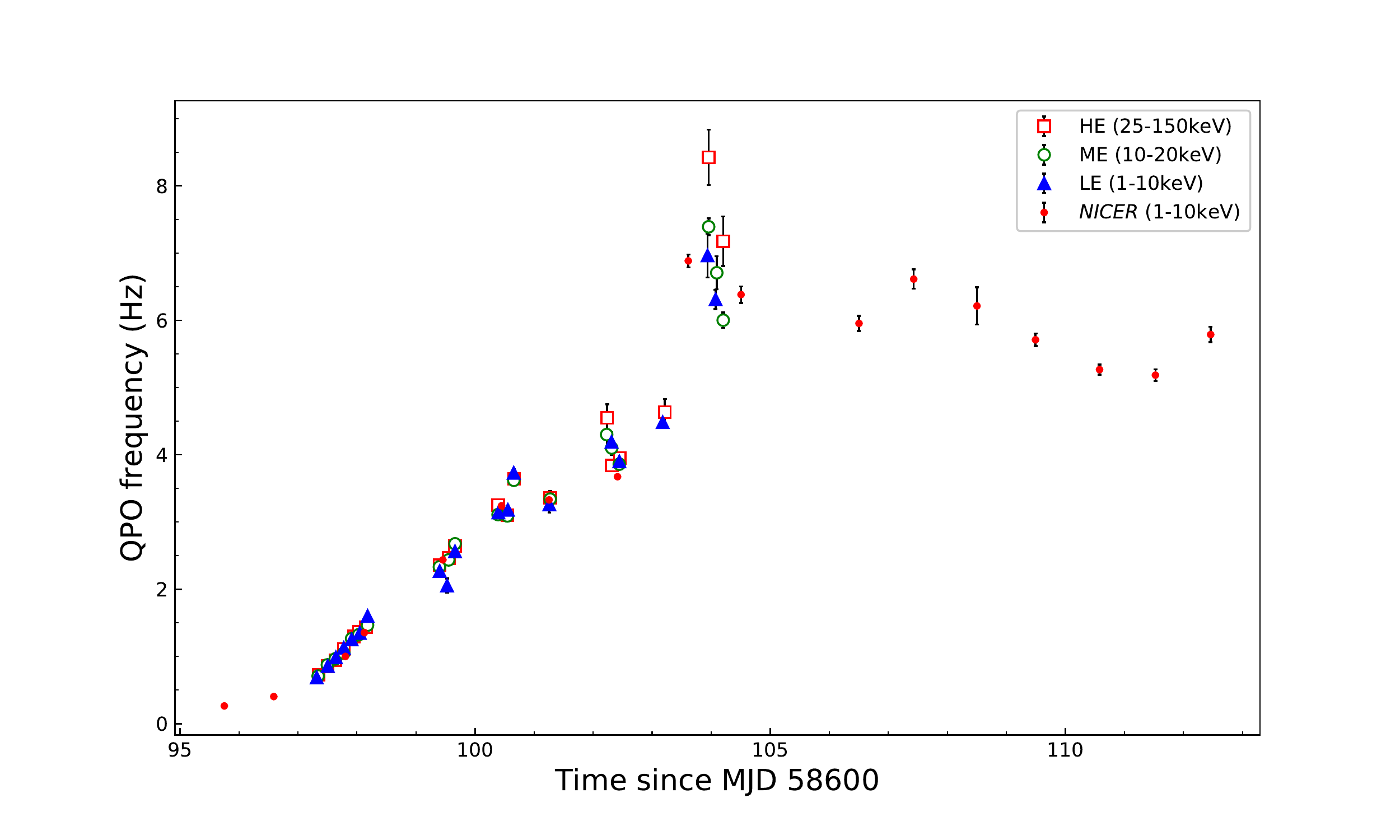}
    \caption{The evolution of QPO frequency observed in this outburst of EXO 1846--031.}
    \label{fig:QPO frequency and LW}
\end{figure}

Figure \ref{fig:QPO frequency and LW} presents QPOs frequency evolution with time for three energy bands of \emph{Insight}--HXMT and 1--10 keV of \emph{NICER}, shown separately in the top and bottom panels. The \emph{Insight}--HXMT data indicates that the frequency of most QPOs shows no energy dependence as seen by LE/ME/HE, and increases monotonically from $0.26\pm0.00$ Hz to $8.42\pm0.41$ Hz. With more {\it NICER} observations, we can see that the frequency of QPOs increases at first and then becomes stable; QPO disappears when the source enters a relatively soft state. During the period when QPOs are observed, the hardness detected ranges from 1.33 to 0.63, with the rms decreases from 23.8\% to 2.66\% respectively.

\begin{figure}
    \centering
    \includegraphics[width=12cm]{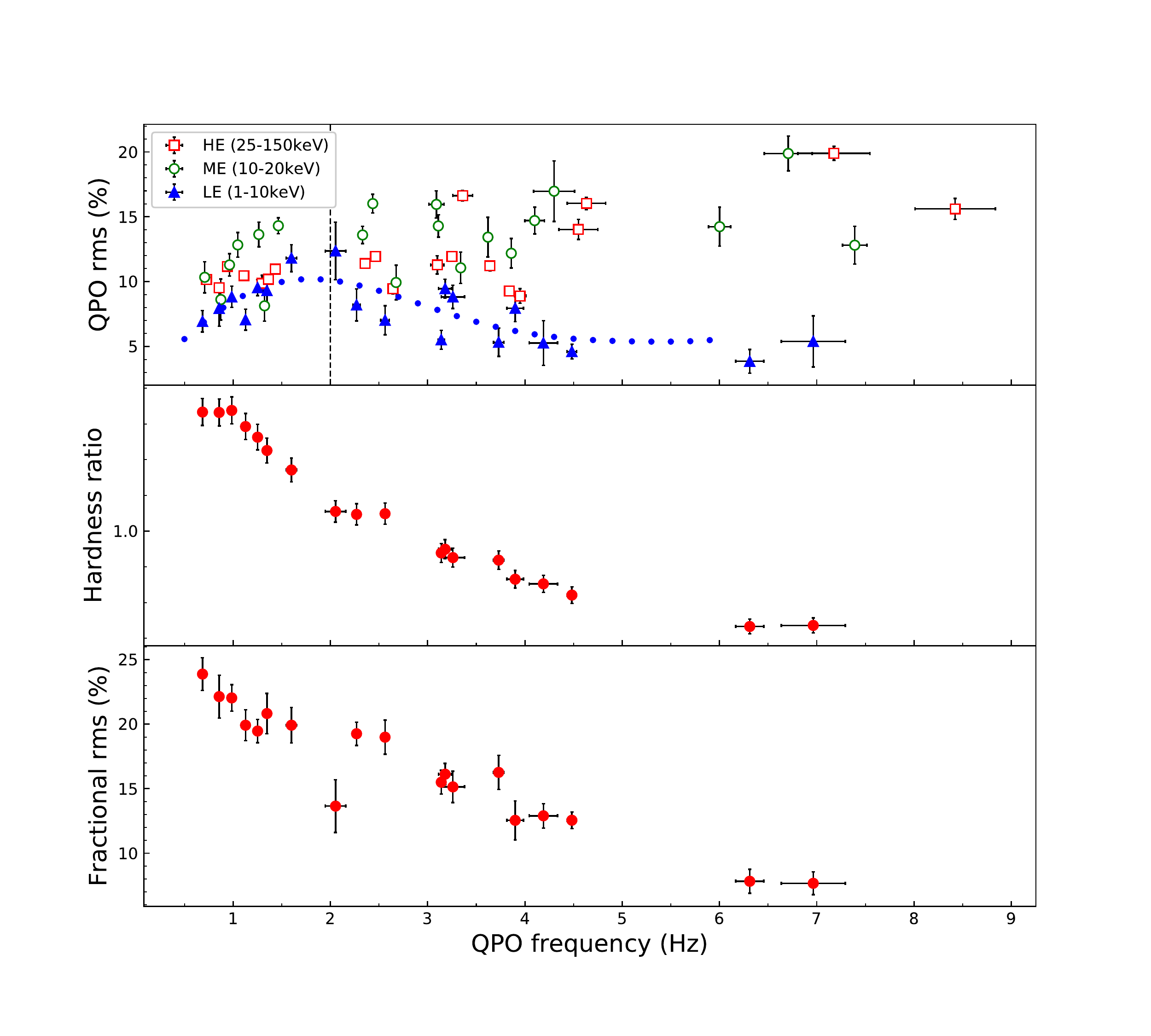}
    \caption{The QPO rms, hardness, and fractional rms as functions of QPO frequency for EXO~1846--031. The QPO rms of \emph{Insight}--HXMT LE/ME/HE are given in 1--10 keV, 10--20 keV, 25--150 keV energy bands, respectively. The hardness is defined as the count rate ratio between 4--10 keV and 2--4 keV.}
    \label{fig:QPO frequency and hardness}
\end{figure}
\begin{figure}
    \centering
    \includegraphics[width=12cm]{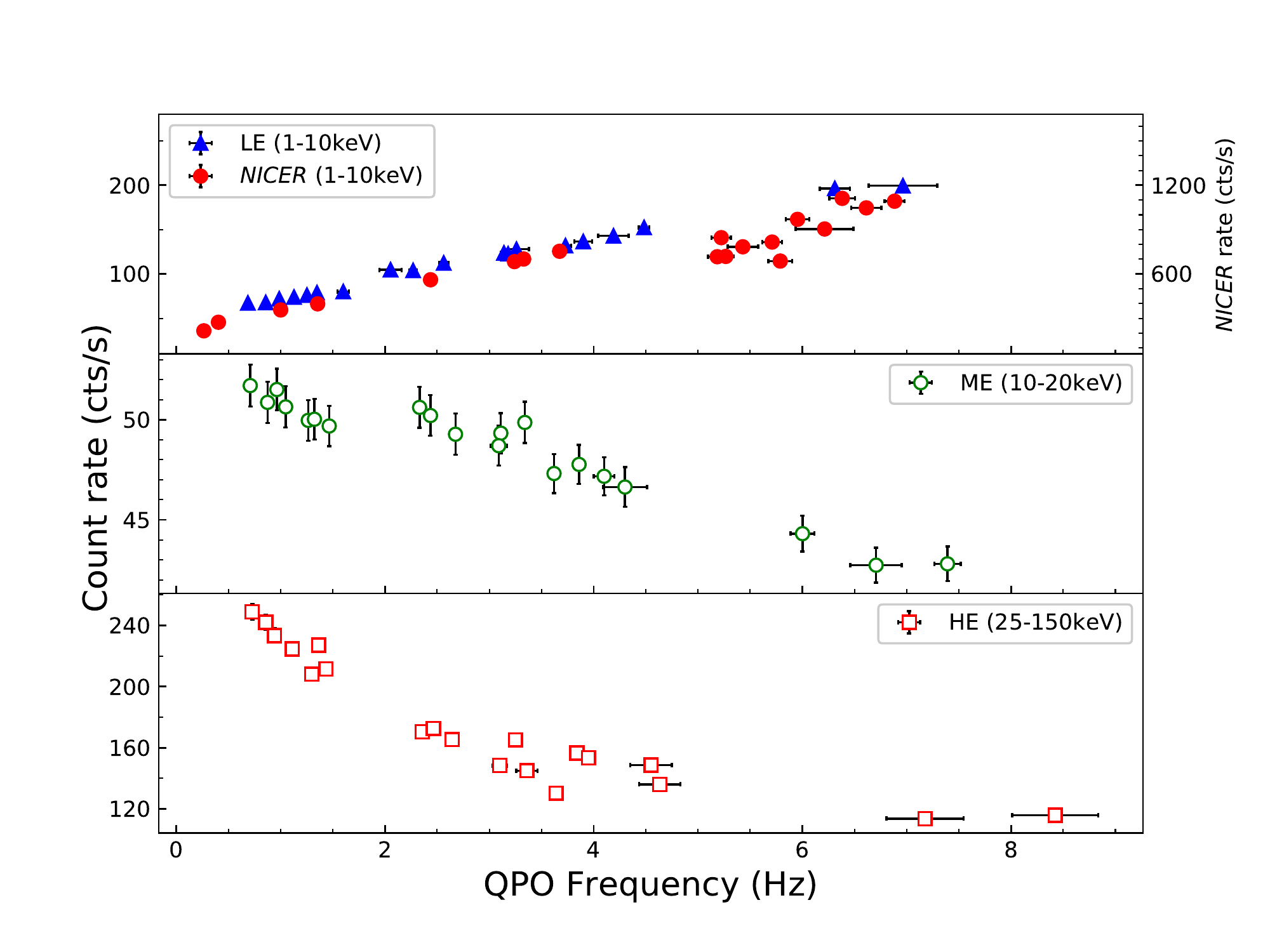}
    \caption{The relationship between QPO frequency and count rate. The frequency and count rate are calculated from \emph{Insight}-HXMT (LE:1--10 keV; ME:10--20 keV; HE:25--150 keV) \& \emph{NICER}.}
    \label{fig:QPO frequency and RATE}
\end{figure}

\label{sec:3.3} 

More intriguingly, as shown in the  top panel of Figure \ref{fig:QPO frequency and hardness}, there is a relationship between fractional rms of QPO and its center frequency. For the QPO detected by LE (energy band: 1--10 keV), the QPOs fractional rms increases from 6.9\% to 12.4\% when the frequency is lower than $\sim$2 Hz, and then decreases, which is similar to the case of GRS~1915+105 \citep{2013MNRAS.434...59Y,2017ApJ...845..143Z}. In the middle panel of Figure \ref{fig:QPO frequency and hardness}, we plot the QPO frequency (LE) as a function of the hardness. As one can see from this figure, the hardness has a downward trend (from $1.31\pm 0.04$ to $0.73\pm 0.02$) with the frequency increasing, while in which we can find there are several flat spots (about 1 Hz, 2 Hz and \textgreater 6 Hz). The relationship between the total fractional rms and frequency is shown in the bottom panel of Figure \ref{fig:QPO frequency and hardness}. We show the QPOs frequency evolution with the count rate in Figure \ref{fig:QPO frequency and RATE}. The QPOs frequency increases with the count rate of LE (1-10 keV), in the meantime the ME (10--20 keV) and HE (25--150 keV) count rate decreases. 

\section{Summary and Discussion}
\label{sec:4}
We have performed a timing analysis and spectral evolution (hardness) study for the 2019 outburst of EXO~1846--031 by using the observations with \emph{Insight}--HXMT, \emph{NICER} and \emph{MAXI}. The HID derived from the \emph{MAXI}/GSC data shows the typical q-shape. Low frequency Type-C QPOs are detected in the early stage of the outburst. The outburst evolution is consistent with the behaviours observed in other typical BHTs, for example, GX 339-4 \citep{2011MNRAS.418.2292M}. Using QPO positions and spectral evolution in HID derived from \emph{Insight}--HXMT/LE, we have identified the different spectral states of the outburst, i.e., LHS, HIMS, SIMS, and HSS. Before MJD 58697, we have found out that the source was in its LHS and then evolved to the HIMS (about MJD 58697 -- MJD 58711). According to our analysis, this outburst appeared in the SIMS for a short time, about a few days. Around MJD 58717, the outburst entered the HSS. Due to the weak X-ray flux when the source returned to its HIMS or LHS, the full deceasing phase of outburst haven't been totally observed by \emph{Insight}--HXMT \& \emph{NICER}. 

The evolution of the total fractional rms values calculated from \emph{Insight}--HXMT/LE observations are similar to those observed in other BHTs \citep{2005A&A...440..207B,2010ApJ...714.1065R,2011MNRAS.410..679M}. During the LHS, the maximum value of rms is 24\% and remaines around 21\%, while rms $\la 21\%$ in HIMS and $5\% \la rms \la 10\%$ in SIMS. In the HSS, the observed rms is between $\sim$ 1.2\% and $\sim$ 10\% with large uncertainties, which seems to be larger than the typical value observed in other BHTs. We obtained a weighted averaged rms value of 3.81\% $\pm$ 2.28\% during HSS, which is consistent with the typical variability amplitude in HSS. A similar behaviour in MAXI J1659--152 is reported by \citet{2011MNRAS.415..292M}, in which they explained it with a high orbital inclination.

Type--C QPOs were both detected during the LHS and HIMS by \emph{Insight}--HXMT and \emph{NICER}. Type–C QPO features are similar to the phenomenon of typical BHTs. The QPO frequency varies from $\sim$0.1 Hz to $\sim$8 Hz with increasing X-ray flux (accretion rate). But no type-B QPO has been detected throughout 2019 outburst in LE and \emph{NICER} observation. The relationships of QPO frequency--hardness and frequency--total fractional rms in the middle and bottom panels of Figure \ref{fig:QPO frequency and hardness} have been reported in other BHTs, such as Swift J1658.2-4242 \citep{2019JHEAp..24...30X}, MAXI J1659--152 \citep{2011MNRAS.415..292M} and MAXI J1820+070 \citep{2019arXiv191207625S}. It is interesting to note that the QPO rms shows different relationship with the QPO frequency at \emph{Insight}-HXMT energy bands in the top panel of Figure \ref{fig:QPO frequency and hardness}, which is similar to Figure 6 in \citet{2020JHEAp..25...29K}. In the low energy band (LE 1--10 keV), the QPO rms first increases, and then decreases at around 2 Hz, which is similar to the results of GRS~1915+105 reported by \cite{2013MNRAS.434...59Y}. Compared with \emph{NICER}, \emph{Insight}--HXMT has a higher effective area at high energies, which can provide the high energy characteristics of QPO. The trend of QPO rms in high (HE 25-150 keV) and medium energy (ME 10-20 keV) continue to rise. The upward trend is slower than the trend in LE energy band before 2 Hz, but after that, the trend remains almost constant. We can see from this figure that at any frequency, the QPO rms at higher energy is greater than that of LE. More detailed discussions of the rms energy spectra of QPO can be seen in GRS 1915+105 \citep{2012Ap&SS.337..137Y}, MAXI~1535--571 \citep{2018ApJ...866..122H}, and H1743--322 \citep{2013MNRAS.433..412L}, in which they considered that the origin of type--C QPOs comes from the non-thermal corona. 

For the LFQPOs, many models have been proposed, most of which can be classified either as geometrical effects -- whereby the shape and/or size of something varies quasi-periodically -- or intrinsic variations -- whereby some fundamental property such as pressure or accretion rate oscillates in a stable geometry (for a review see, e.g. \citealt{2020arXiv200108758I}). In this work, we primarily incline to use the Lense-Thirring precession model \citep{1998ApJ...492L..59S,2009MNRAS.397L.101I} to explain the observational phenomena, while also discussing our results in the framework of AEI model \citep{1999A&A...349.1003T}.

In the truncated disk model \citep{1997ApJ...489..865E}, the thin disc truncates at some radius, and is replaced by a hot inner flow (corona) which is thought to produce the non-thermal emission. If the black hole spin misaligns with orbits spin, the inner flow will undergo Lense--Thirring precession, which modulates the flux mainly by the combination of Doppler boosting and projected area effects. According to the definition, $rms=\Delta F/(TC+PL)$ (where $\Delta F$ is the QPO amplitude, \emph{TC} is the thermal emission from the thin disk and \emph{PL} is the non--thermal emission flux produced by the inner hot flow), we know that the rms of QPO is affected by both thermal and non-thermal components. During the state transition, the truncation radius moves in and the inner hot flow contracts to a more central position and becomes more compact   \citep{2019Natur.565..198K}, leading to softening of the spectra and increasing of the QPO frequency. When the QPO frequency is lower than $\sim$2 Hz (about MJD 58698), the energy spectrum is dominated by non--thermal component (Ren et al., manuscript in preparation). As the mass accretion rate increases, the optical depth of the inner flow increases, thus giving stronger modulation \citep{2009MNRAS.397L.101I,2006ApJ...642..420S}. In addition, the reduction of non-thermal components also leads to the increase of rms of QPO. On the other hand, when the QPO frequency is above $\sim$2 Hz, as the inner disc approaches the black hole, soft photons from the disk continue to cool hot electrons in the inner hot flow, making the size of inner hot flow decreases, and thermal component gradually begins to play a major role in the low energy band. As shown in Figure \ref{fig:LC}, the count rate of LE continues to increase after $\sim$2 Hz, which leads to the decrease of rms of QPO in the lower energy band. However, the emission of HE and ME comes from non--thermal flow. Therefore, the rms of QPO presents a flattening or even a slightly rising trend later in higher energy.

In addition to the precession models, there is another class to discuss the origin of QPO based on various instabilities occurring in either the disk or the corona. Among these models, we intend to use the AEI model to explain our results. The AEI model \citep{2001AIPC..587..131V,2002A&A...387..487R,2002A&A...387..497V} suggests that the instability of global spiral on magnetized accretion disk may be the origin of QPO. The rotation frequency of the quasi-steady spiral is the orbital frequency at the co-rotation radius ($r_{\rm c}$), which is a few times of the orbital frequency at the inner edge of the disk ($r_{\rm in}$). Based on this, \citet{2009ApJ...694L.132M} discussed the relationship between QPO frequency and inner disk radius. Using the AEI model, \citet{2017ApJ...834..188V} explained the QPO rms-frequency relationship in the 1998-99 outburst of XTE J1550-564, which is similar to the results of this article, but the maximum rms at $\sim$1.5 Hz. They suggest that both the strength of the instability and the unmodulated flux emitted between $r_{\rm in}$ and $r_{\rm c}$ should affect the QPO rms. Thus different outbursts from the same sources or between sources will have a similar increase and then decrease, but with a different position for its maximum which depends on the local condition in the disk (such as its density, magnetization, etc.).

Recently, \citet{2015MNRAS.447.2059M} reported such an inclination dependence of QPO properties with statistical relation between the QPO rms and its frequency. They find that the type--C QPO shows a systematically larger absolute variability amplitude in edge--on sources, consistent with Lense--Thirring precession as its origin. They also find the rms-frequency relation has a maximum value at around 2 Hz (see Figure 1 in \citealt{2015MNRAS.447.2059M}). So, for the phenomenon at 1--10 keV, one speculation is that the rms of QPOs depends on the orbital inclination. This result might suggest that EXO~1348-031 is a high inclination system. However, it is unclear why the above 2 Hz feature is observed in high inclination systems.

\begin{acknowledgements}
This work made use of the data from the HXMT mission, a project funded by China National Space Administration (CNSA) and the Chinese Academy of Sciences (CAS). This research has applied \emph{MAXI} data provided by RIKEN, JAXA, and the \emph{MAXI} team. This work also has acquired the data from the High Energy Astrophysics Science Archive Research Center Online Service, provided by the NASA/Goddard Space Flight Center.This work is supported by the National Key R\&D Program of China (2016YFA0400800) and the National Natural Science Foundation of China (NSFC) under grants 11673023, U1838201, U1838115, U1838111, U1838202, 11733009, U1838108.

\end{acknowledgements}

\bibliographystyle{raa}
\appendix
\section{}
\begin{table*}
    \centering
    \begin{tabular}{c c c c c c}
     \hline
       ObsId &  MJD & exposure(s) & frequency(Hz) & Q & rms(\%) \\
     \hline
     P021405000101 & 58697.3 & 1074 & $0.69\pm0.01$& $10.5\pm4.9$ & $6.93\pm0.82$ \\
    
     P021405000102 & 58697.5 & 705 & $0.85\pm0.01$ & $10.1\pm6.9$ & $7.93\pm1.38$ \\
     
     P021405000103 & 58697.6 & 1656 & $0.99\pm0.01$ & $7.1\pm2.1$ & $0.82\pm0.82$ \\
     
     P021405000104 & 58697.8 & 1285 & $1.13\pm0.02$ & $9.1\pm6.9$ & $7.93\pm1.38$ \\
     
     P021405000105 & 58697.9 & 2274 & $1.25\pm0.01$ & $6.1\pm1.2$ & $9.53\pm0.61$ \\
     
     P021405000106 & 58698.0 & 598 & $1.34\pm0.01$ & $12.3\pm4.4$ & $9.31\pm0.95$ \\
     
     P021405000107 & 58698.2 & 859 & $1.60\pm0.05$ & $2.7\pm0.8$ & $11.81\pm1.04$ \\
     
     P021405000201 & 58699.4 & 1232 & $2.26\pm0.06$ & $6.1\pm2.3$ & $8.20\pm1.23$ \\
     
     P021405000202 & 58699.5 & 418 & $2.05\pm0.11$ & $12.1\pm5.5$ & $2.03\pm0.88$ \\
     
     P021405000203 & 58699.7 & 478 & $2.56\pm0.04$ & $10.1\pm5.5$ & $7.02\pm1.12$ \\
     
     P021405000301 & 58700.4 & 1234 & $3.14\pm0.03$ & $15.1\pm6.4$ & $5.51\pm0.73$ \\
     
     P021405000302 & 58700.6 & 1404 & $3.18\pm0.07$ & $3.5\pm0.8$ & $9.46\pm0.73$ \\
     
     P021405000303 & 58700.7 & 478 & $3.73\pm0.05$ & $16.1\pm11.1$ & $5.32\pm1.08$ \\
     
     P021405000401 & 58701.2 & 718 & $3.26\pm0.12$ & $3.1\pm1.1$ & $8.82\pm0.90$ \\
     
     P021405000502 & 58702.3 & 1250 & $4.19\pm0.15$ & $5.7\pm4.3$ & $5.26\pm1.72$  \\
     
     P021405000503 & 58702.4 & 538 & $3.90\pm0.09$ & $6.2\pm2.6$ & $7.94\pm1.04$ \\
     
     P021405000601 & 58703.2 & 2506 & $4.48\pm0.05$ & $13.9\pm5.1$ & $4.61\pm0.56$ \\
     
     P021405000701 & 58703.9 & 1915 & $6.69\pm0.33$ & $3.5\pm2.5$ & $5.39\pm1.97$ \\
     
     P021405000702 & 58704.1 & 1795 & $6.31\pm0.14$ & $9.1\pm6.1$ & $3.86\pm0.91$ \\
     \hline
     
    \end{tabular}
    \caption{Fit parameters of QPOs observations of EXO~1846--031 extracted from \emph{Insight}--HXMT LE (1--10keV). The centroid frequency, the quality factor Q, fraction rms are shown in this table.}
    \label{tab:LE QPOs}
\end{table*}

\begin{table*}
    \centering
    \begin{tabular}{c c c c c c}
     \hline
       ObsId & MJD  & exposure(s) & frequency(Hz) & Q & rms(\%) \\
     \hline
     P021405000101 & 58697.3 & 3490 & $0.71\pm0.01$ & $4.9\pm1.7$ & $10.33\pm1.19$ \\
     
     P021405000102 & 58697.5 & 2397 & $0.87\pm0.02$ & $6.6\pm3.8$ & $8.62\pm1.58$ \\
    
     P021405000103 & 58697.6 & 2355 & $0.96\pm0.01$ & $5.9\pm1.6$ & $11.29\pm0.86$ \\
     
     P021405000104 & 58697.8 & 1949 & $1.05\pm0.03$	& $3.3\pm0.9$	& $12.83\pm0.95$ \\
     
     P021405000105 & 58697.9 & 2041 &$1.26\pm0.02$ & $5.3\pm1.2$ & $13.63\pm0.95$ \\
     
     P021405000106 & 58698.0 & 2833 &$1.32\pm0.01$ & $21.1\pm10.4$ & $8.13\pm1.18$ \\
     
     P021405000107 & 58698.2 & 2999 & $1.47\pm0.02$	& $5.1\pm0.9$ & $14.31\pm0.62$ \\
     
     P021405000201 & 58699.4 & 3429 & $2.33\pm0.02$ & $7.4\pm1.3$	& $13.60\pm0.67$ \\
     
     P021405000202 & 58699.5 & 2267 & $2.44\pm0.03$	& $5.4\pm0.9$	& $16.02\pm0.72$ \\
     
     P021405000203 & 58699.7 & 578 & $2.68\pm0.03$ & $32.2\pm31.3$ & $9.93\pm1.33$ \\
     
     P021405000301 & 58700.4 & 3378 & $3.11\pm0.04$ & $5.1\pm1.1$ & $14.29\pm0.86$ \\
     
     P021405000302 & 58700.6 & 1936 & $3.09\pm0.08$ & $3.4\pm0.9$	& $15.96\pm1.05$ \\
     
     P021405000303 & 58700.7 & 929 & $3.62\pm0.04$ & $10.6\pm4.8$& $13.43\pm1.53$ \\
     
     P021405000401 & 58701.2 & 795 & $3.34\pm0.02$& $27.8\pm19.9$	& $11.06\pm1.21$ \\
     
     P021405000501 & 58702.2 & 439 & $4.30\pm0.21$& $3.9\pm2.1$ & $16.97\pm2.33$ \\
     
     P021405000502 & 58702.3 & 2939 & $4.10\pm0.01$ & $3.8\pm1.1$ & $14.71\pm1.03$ \\
     
     P021405000503 & 58702.4 & 976 & $3.86\pm0.04$ & $18.4\pm7.2$ & $12.19\pm1.14$ \\
     
     P021405000701 & 58703.9 & 2537 & $7.39\pm0.13$ & $8.5\pm3.8$ & $12.81\pm1.46$ \\
     
     P021405000702 & 58704.1 & 2321 & $6.71\pm0.25$ & $2.8\pm0.8$	& $19.86\pm1.33$ \\
     
     P021405000703 & 58704.2 & 2598 & $6.00\pm0.11$	& $6.5\pm2.6$	& $14.23\pm1.50$ \\

     \hline

    \end{tabular}
    \caption{Fit parameters of QPOs observations of EXO~1846--031 extracted from \emph{Insight}--HXMT ME (10--20keV). The centroid frequency, the quality factor Q, fraction rms are shown in this table.}
    \label{tab:ME QPOs}
\end{table*}

\begin{table*}
    \centering
    \begin{tabular}{c c c c c c}
     \hline
       ObsId & MJD  & exposure(s) & frequency(Hz) & Q & rms(\%) \\
     \hline
     P021405000101 & 58697.3 & 3707 & $0.73\pm	0.01$ & $5.1\pm0.9$ & $10.16\pm0.25$ \\
     
     P021405000102 & 58697.5 & 3969 & $0.86\pm	0.01$ & $6.1\pm0.9$ & $9.52\pm0.19$ \\
    
     P021405000103 & 58697.6 & 3730 & $0.94\pm0.01$ & $4.6\pm0.7$ & $11.17\pm0.25$ \\
     
     P021405000104 & 58697.8 & 2204 & $1.11\pm0.01$ & $5.2\pm0.9$	& $10.46\pm0.26$ \\
     
     P021405000105 & 58697.9 & 653 & $1.29\pm0.01$ & $13.5\pm8.1$ & $9.89\pm0.62$ \\
    
     P021405000106 & 58698.0 & 4004 & $1.36\pm0.01$ & $7.1\pm0.9$ & $10.18\pm0.18$ \\
     
     P021405000107 & 58698.2 & 1426 & $1.43\pm0.01$& $8.6\pm1.7$ & $10.97\pm0.25$ \\
     
     P021405000201 & 58699.4 & 4964 & $2.36\pm0.02$ & $6.8\pm1.1$	& $11.40\pm0.19$ \\
     
     P021405000202 & 58699.5 & 3845 & $2.46\pm0.03$ & $5.8\pm1.1$ & $11.94\pm0.24$ \\
     
     P021405000203 & 58699.7 & 907 & $2.64\pm0.03$ & $15.9\pm6.6$ & $9.46\pm0.39$ \\
     
     P021405000301 & 58700.4 & 4873 & $3.25\pm0.03$ & $6.3\pm1.3$	& $11.94\pm0.29$ \\
     
     P021405000302 & 58700.6 & 3542 & $3.10\pm0.07$ & $4.8\pm2.1$& $11.29\pm0.70$ \\
     
     P021405000303 & 58700.7 & 1690 & $3.64\pm0.04$ & $14.7\pm6.3$ & $11.21\pm0.38$\\
     
     P021405000401 & 58701.2 & 1090 & $3.36\pm0.10$ & $3.6\pm1.1$ & $16.63\pm0.41$ \\
     
     P021405000501 & 58702.2 & 437 & $4.55\pm0.20$ & $4.8\pm2.8$ & $14.02\pm0.77$ \\
     
     P021405000502 & 58702.3 & 4176 & $3.84\pm0.03$ & $14.6\pm3.6$ & $9.27\pm0.22$ \\
     
     P021405000503 & 58702.4 & 1670&$	3.95\pm	0.06$ & $11.9\pm7.1$ & $8.89\pm0.56$ \\
     
     P021405000601 & 58703.2 & 1570 & $4.63\pm0.20$ & $2.9\pm1.1$	& $16.03\pm0.48$ \\
     
     P021405000701 & 58703.9 & 1570 & $8.42\pm0.41$& $2.8\pm1.5$ & $15.59\pm0.81$ \\
     
     P021405000702 & 58704.1 & 173 & $6.71\pm0.25$& $2.8\pm	0.8$ & $19.86\pm1.33$ \\
     
     P021405000703 & 58704.2 & 2135 & $7.18\pm0.37$ & $2.4\pm0.9$ & $19.90\pm0.53$ \\

     \hline

    \end{tabular}
    \caption{Fit parameters of QPOs observations of EXO~1846--031 extracted from \emph{Insight}--HXMT HE (25--150keV). The centroid frequency, the quality factor Q, fraction rms are shown in this table.}
    \label{tab:HE QPOs}
\end{table*}

\begin{table*}
    \centering
    \begin{tabular}{c c c c c c}
     \hline
       ObsId & MJD  & exposure(s) & frequency(Hz) & Q & rms(\%) \\
     \hline
     2200760101 & 58695.8 &	3054 & $0.26\pm0.01 $ & $7.9\pm2.3$ & $5.81\pm0.54 $ \\
     
     2200760102 & 58696.6 &	5658 & $0.40\pm0.01$ & $10.7\pm4.7$ & $6.32\pm0.72 $ \\
     
     2200760103 & 58697.8 &	3054 & $1.00\pm0.01$ & $2.9\pm	0.4 $ &	$10.48\pm0.59$ \\
    
     2200760104 	&	58698.1 	&	5658	& $	1.35 	\pm	0.02 	$&	$3.1 	\pm	0.5 	$ &	$10.61 	\pm	0.78 	$\\
     
     2200760105 &58699.5 & 3054 & $2.44\pm0.03$ &	$2.8\pm0.8$ & $8.25\pm1.36 $\\
    
     2200760106 & 58700.4 &	5658 & $3.24\pm0.02$ &$5.8\pm	0.8$ &$5.86\pm0.37$\\
     
     2200760107 & 58701.3 &  5600 & $3.33\pm0.02$ & $8.9\pm	1.5$ & $6.02\pm0.39$\\
     
     2200760108 & 58702.4 &	5658 & $ 3.67\pm0.02$ &	$10.8\pm	1.9$ & $4.96\pm0.31$\\
    
     2200760109 & 58703.6 &	1165 & $6.88\pm0.10$ & $6.5\pm2.1$ & $ 2.34 \pm0.29$\\
     
     2200760110 & 58704.5 &	2562 & $6.38\pm0.12$ & $3.2\pm0.7$ & $3.01\pm0.31$\\
    
     2200760112 & 58706.5 &	1130 & $5.95\pm0.11$ & $5.5\pm2.1$ & $2.47\pm0.39$\\
     
     2200760113 & 58707.4 & 3564 & $6.61\pm0.15$ & $8.7\pm	5.3$ & $1.50\pm0.34$\\
     
     2200760114 & 58708.5 &	912	& $6.21\pm0.28$ & $2.7\pm0.9 	$ &	$2.99\pm0.87$\\
     
     2200760115	& 58709.5 &	927	& $	5.71\pm0.10$ & $3.5\pm0.9$ & $3.13\pm0.48$\\
     
     2200760116 & 58710.6 &	3293 & $5.27\pm0.08$ & $3.6\pm0.6$ & $3.97\pm0.29$\\
     
     2200760117 & 58711.5 &	4629 & $5.18\pm0.09$ & $4.4\pm1.6$ & $2.65\pm0.56$\\
     
     2200760118 & 58712.5 &	3771 & $5.79\pm0.12$ & $5.2\pm	2.3$ & $2.52\pm0.55$\\
     
     2200760119 & 58713.5 & 2749 & $5.22\pm0.09$ &	$5.8\pm	1.9$ & $2.31\pm0.27$ \\
     2200760120 & 58714.5 &	3341 & $5.43\pm0.15$ & $6.9\pm4.2$ & $1.66\pm0.38$  \\
    \hline

    \end{tabular}
    \caption{Fit parameters of QPOs observations of EXO~1846--031 extracted from \emph{NICER} (1--10 keV). The centroid frequency, the quality factor Q, fraction rms are shown in this table.}
    \label{tab:NICER QPOs}
\end{table*}

\end{document}